\documentclass{amsart}
\usepackage[frenchb,english]{babel}
\usepackage{graphicx}
\usepackage{subcaption}
\usepackage{mathabx}

\usepackage{stmaryrd}

\usepackage{natbib}

\usepackage[margin=1.2in]{geometry}
\usepackage{multirow}

\theoremstyle{definition}

\theoremstyle{remark}

\numberwithin{equation}{section}
\newcommand{\norm}[1]{\left\lVert#1\right\rVert}
\newcommand{\ie}{\textit{i.e.,\ }}
\newcommand{\eg}{\textit{e.g.,\ }}

\usepackage{rotating}
\usepackage{breakcites}
\usepackage{lineno}
\usepackage[colorlinks,citecolor=blue]{hyperref}
\begin{document}

\title{Drift and its mediation in terrestrial orbits}
%    Remove any unused author tags.

%    author one information
%    author one information
\author{J\'er\^ome Daquin}
\address{
Department of Mathematics ''Tullio Levi-Civita"
Via Trieste, 63, 35131 Padova, Italy.
}
\address{School of Science, RMIT University, RMIT City Campus, GPO Box 2476V, Melbourne Victoria 3001, Australia.}
\address{Space Environment Research Centre (SERC) Limited, Mount Stromlo Observatory, Canberra, Australian Capital Territory, Australia.}
%\curraddr{}
\email{daquin@math.unipd.it}
%\thanks{Special thanks are due to}
%    author two information

\author{Ioannis Gkolias}
\address{Department of Aerospace Science and Technology, Politecnico di Milano, Milan, Italy.}
%\email{}

\author{Aaron J. Rosengren}
\address{Aerospace and Mechanical Engineering, University of Arizona, Tuscon, Arizona, United States.}
%\curraddr{}
%\email{}
%\thanks{}
%\keywords{One, two, three.}
%\subjclass[2010]{Primary }

\date{\today}

%\maketitle
%\subjclass[2010]{Primary }
%\keywords{medium Earth orbits - numerical integration - keyword 3}
\date{\today}

\dedicatory{}
\begin{abstract}
%%% Leave the Abstraceyt empty if your article does not require one, please see the Summary Table for full details.
The slow deformation of terrestrial orbits in the medium range, subject to lunisolar resonances, is well approximated by a family of Hamiltonian flow with $2.5$ degree-of-freedom.  The action variables of the system may experience chaotic variations and large drift that we may quantify. Using variational chaos indicators, we compute high-resolution portraits of the action space. Such refined meshes allow to reveal the existence of tori and structures filling chaotic regions. Our elaborate computations allow us to isolate precise initial conditions near specific zones of interest  and  study their asymptotic behaviour in time.  Borrowing classical techniques of phase-space visualisation, we highlight how the
drift is mediated by the complement of the numerically detected  KAM tori.\\
 
\end{abstract}

\maketitle 
\tableofcontents
%----------------------------------------------------
\section{Introduction}
%----------------------------------------------------
Various groups of scientists have become enchanted anew by the lunisolar resonances affecting the dynamics of terrestrial orbits. The study of them and the resurgence of their significance has not been visible since the notorious and colossal triptych of \cite{sBr99,sBr01-2,sBr01}. Later rebranded by \cite{aRo08} in the context of the \textit{medium-Earth orbits} (MEOs), the study of their long-term dynamics, and in particular their eccentricity growths in the \textit{elliptic domain} \citep{bBo09}, represent current deep motivations for the community. In our opinion, the most complete and up-to-date panorama of the literature is excellently presented by \cite{rAr18}. The existence of such a condensation allows us to adopt here a rather direct {\color{black} style in} this present contribution. We are particularly interested by questions related to the stability of orbits. Based on the divergence of nearby trajectories, the existence of a mixed phase space where there is a cohabitation of stable and chaotic components has been recently pictured \citep{jDa16,aCe16,aCe16-siam,iGk16,aCe17} and partially  explained applying Chirikov's resonances overlap criterion \citep{bCh79}. The Hamiltonian flow obtained under the simplest 
assumptions for the disturbing effects of the perturbers (\ie a development restricted to its lowest order and averaged over fast variables), Moon and Sun, encapsulates all the details of the dynamics {\color{black} in which} we are interested \citep{jDa16}. In particular, the Hamiltonian {\color{black} possesses} $2$ {\color{black} degrees}-of-freedom (DOF) and depends periodically  on the time $t$ (see \cite{aCe17} for omitted details). 
We recall in the next section how the Hamiltonian 

\begin{equation}\label{Eq:Hamiltonian}
\begin{aligned}
   	\mathcal{H} \colon &D \times \mathbb{T}^{2} \times \mathbb{T} \to \mathbb{R}, \, (x,y,t) \mapsto \mathcal{H}(x,y,t)=h_{0}(x) + \varepsilon h_{1}(x,y,t)
 \end{aligned}
\end{equation}  

with $h_{1}(x,y,t)=\sum_{m \in \mathcal{A} \subset \mathbb{Z}_{\star}^{3}} h_{m}(x) \cos (m\cdot (y,t) + \phi_{m})$
is obtained. The form of Eq.\,(\ref{Eq:Hamiltonian}) is the standard form of a nearly-integrable problem written in action-angles variables. The non-linearity parameter $\varepsilon$ belongs to a certain subinterval $\mathcal{I}$ of $\mathbb{R}_{+}$ and is function of the semi-major axis, which is a first integral in the secular approximation. The functions $\{h_{m}\}_{m \in \mathcal{A}}$ are real valued functions of the sole action $x \in D \subset \mathbb{R}^{2}$ (and some constant physical parameters),   
the $\phi_{m}$ are {\color{black}phase} terms.
When $\varepsilon$ sweeps $\mathcal{I}$, a transition from a globally ordered {\color{black} phase space} to a mixed phase space is known to exist. It turns out that the {\color{black}presence} of a chaotic regime for large {\color{black}values} of $\varepsilon$, say for $\varepsilon$ close to $\max \mathcal{I}$, corresponds to the range of semi-major {\color{black}axes} where the navigation satellites are located. The {\color{black}occurrence} of the two apparent antonyms, `how awkward it is'\footnote{
The  Lyapunov times $\tau_{\mathcal{L}}$, which dynamically speaking constitute the  barriers of predictability, are on the order {\color{black}of decades} \citep{jDa16}.} and `how useful and fruitful it can be'\footnote{
There is a birth of a {\color{black}new ideology} to remedy the space-debris problem, based on a `judicious' use of the instabilities to define re-entry orbits and navigate the {\color{black}phase space.}}, crystallizes assuredly the challenges, implications and beauty of the dynamical and {\color{black}engineering} problems we face.
{\color{black}
Gravitational problems are  kaleidoscopes of pure and applied science. Our Solar System has been the source and receptacle of many theoretical and practical dynamical facets and aspects (KAM theory, hyperbolic dynamics, shadowing theory, numerical analysis, phase visualisation techniques). Spaceflight dynamics is not excluded and has gained from this rich heritage \citep{eBo95,wKo00,ePe10}. We cannot take a definitive position on the space-debris mitigation via `chaos targeting' and transfers in phase space, nevertheless, let us underline that the concept embraces the continuous necessary exchanges between (technological and scientific) communities.  
}

In this paper, we depart from former goals where the main impetus was the explanation of the mechanisms supporting the apparition of chaos. Instead, we focus rather on i) {\color{black}the} physical consequences in terms of transport in the phase space and ii) on the visualisation of these excursions \textit{via}  double sections in the action-like {\color{black}phase space}.  The techniques we used have been extensively employed  
in Dynamical Astronomy and overall in the context of the dynamics of quasi-integrable Hamiltonian systems and symplectic discrete maps {\color{black}\citep[confer][just to name a few]{eLe03,nTo08,pCi08,riPa15,eLe16}}. 
To achieve our tasks, we provide a cartographic view of the prograde \textit{and} retrograde region in {\color{black}Section~\ref{subsec:PSV}, based on {\color{black}a} lighting-fast \textit{ad-hoc} secular model that we recall in Section~\ref{SubSec:EDOs}.} The fine resolutions of the meshes used to discretize the domains $D$ allow {\color{black}for highly detailed views of the phase space}. 
We then focus on the computation of diameters-like quantities to relate the degree of hyperbolicity (a local property) with a more practical transport-like index (a global property). Thanks to our resolved grids, 
precise initial conditions (ICs) {\color{black}can be extracted, which lie near specific structures of interest, in} particular where large diameters are expected. Once {\color{black}obtained,} we proceed to their asymptotic analysis (in time) {\color{black}using ensemble orbit propagation (Section~\ref{Sec:DriftTransport})}. We close with {\color{black}Section~\ref{sec:discussion}} where we summarise our contributions and discuss an open problem that {\color{black}inspires our future} efforts.

%------------------------------------------------------------
\section{The model} 
%------------------------------------------------------------
We recall briefly, for the sake of completeness, under which hypotheses the $2.5$-DOF Hamiltonian is obtained. After the presentation of the model, we present to the newcomers a few facets of the resonant aspects. 

%------------------------------------------------------------
\subsection{Derivation of the Hamiltonian model}\label{subset:model} 
%------------------------------------------------------------
Numerical evidence has shown that, for the range of the treated perturbation $\mathcal{I}$ ({\color{black}recall $\mathcal{I}\simeq[2.2 r_{\Earth},4.65 r_{\Earth}]$}), refinements of the gravitational potentials beyond the quadrupolar level are not necessary to capture details of the {\color{black}{\it global}} dynamics we are interested in, even on {\color{black}long timescales} \citep{jDa16}. It means that when the potentials of the Earth \textit{and} those of the external bodies, Moon and Sun, are developed using Legendre expansions, terms with $l > 2$ are disregarded.  By recognising the timescales of the dynamics, further simplifications are even possible to get a {\color{black}more} pertinent analytical model (and numerical as well; see also Section~\ref{SubSec:EDOs}). Based on the Lagrangian averaging principle \citep{eGr65,iMi67,eGh07}, the potentials are averaged over the mean anomaly of the {\color{black}test particle} $\ell$ and those of the {\color{black}third bodies}\footnote{{\color{black}
In the following, we use the subscripts $\bullet_\Earth$, $\bullet_\Moon$, $\bullet_\Sun$ to denote {\color{black}parameters} referring
to the Earth, Moon and {\color{black}Sun, respectively}.
}},
$\ell_{\Sun}$ and $\ell_{\Moon}$. 

For an oblate Earth, we {\color{black}recall} the classical {\color{black}averaged} potential 
\begin{align}
	\mathcal{H}_{J_{2}}(G,H)= \alpha_{J_{2}}  \big(G^{-3}-3G^{-5}H^{2}\big)
\end{align}
expression, with $\alpha_{J_{2}}=J_{2}r_{\Earth}^{2}\mu_{\Earth}^{4}/4L^{3} \in \mathbb{R}$. Here $(G,H)$ denotes the second and third {\color{black}variables} {\color{black}of the Delaunay actions $(L,G,H)$}, {\color{black} $\mu_{\Earth}$ denotes the (Earth's) gravitational parameter}. The canonically conjugated vector of angles is classically denoted $(\ell,g,h)$.
Omitting details that might be found in \cite{aCe16-siam,aCe17}, 
the disturbing function of the Sun's attraction, $\mathcal{R}_{\Sun}$, reads as
\begin{align}\label{Eq:R-Sun}
	\mathcal{R}_{\Sun}(G,H,g,h)= \sum_{m=0}^{2} \sum_{p=0}^{2} \alpha_{\Sun} \textrm{s}_{m}
	F_{2,m,p}(i)F_{2,m,1}(i_{\Sun})H_{2,p,2p-2}(e)\cos\big(f_{m,p}(g,h)\big)
\end{align}
with
\begin{align}
f_{m,p}: \, \mathbb{T}^{2} \to \mathbb{T}, \,  (g,h)\mapsto(2-2p)g + m(h-h_{\Sun}).
\end{align}
The scalar $\alpha_{\Sun}=\mu_{\Sun} \Big(\frac{a^{2}}{a_{\Sun}^{3}}\Big)(1-e_{\Sun}^{2})^{-3/2}$ has a constant magnitude of  $\sim 3.96 \times 10^{-14}$ in the international system of units.
The coefficients $\textrm{s}_{m}$ are defined as $\textrm{s}_{m}=K_{m} (2-m)!/(2+m)!$. 
{\color{black} The functions $F_{2,m,p}(\bullet)$ refers to Kaula's inclination function \citep{wKa66} and $H_{2,p,2p-2}(e)$ is related to {\color{black}the Hansen coefficients}.}
For the disturbing function of the Moon, the following formula holds true 
\begin{align}\label{Eq:R-Moon}
	\mathcal{R}_{\Moon}(G,H,g,h,h_{\Moon})= 
	\sum_{m=0}^{2} \sum_{p=0}^{2} \sum_{s=0}^{2}
	\alpha_{\Moon}
	\textrm{m}_{m,s}
	F_{2,m,p}(i)F_{2,s,1}(i_{\Moon})H_{2,p,2p-2}(e)\\ \notag
	\times 
	\big(
	U_{2}^{m,-s}\cos\big(g_{m,p,s}(g,h)\big)
	+
        U_{2}^{m,s}\cos\big(h_{m,p,s}(g,h)\big)
	\big)
\end{align}
with
\begin{align}
g_{m,p,s}: \, \mathbb{T}^{2} \to \mathbb{T}, \,  (g,h)\mapsto(2-2p)g + mh +s h_{\Moon} + s \frac{\pi}{2}-y_{s}\pi, \\ 
h_{m,p,s}: \, \mathbb{T}^{2} \to \mathbb{T}, \,  (g,h)\mapsto(2-2p)g + mh -s h_{\Moon} - s \frac{\pi}{2}-y_{s}\pi. 
\end{align}
{\color{black}The expressions $U_{2}^{m,\pm s}$ are function of the obliquity of the {\color{black} eccliptic, and are present due to a rotation of the spherical harmonics needed in this mixed-reference frame formalism}.}
Note that the size of the coefficient 
$\alpha_{\Moon} = \frac{\mu_{\Moon}}{2} \Big(\frac{a^{2}}{a_{\Moon}^{3}}\Big)(1-e_{\Moon}^{2})^{-3/2} \sim 4.32 \times 10^{-14}$ is close to $\alpha_{\Sun}$ (The ratio $\alpha_{\Moon}/\alpha_{\Sun} \sim 1.09$). The coefficients $\textrm{m}_{m,s}$ are defined as $\textrm{m}_{m,s}=(-1)^{[m/2]}K_{m}K_{s} (2-s)!/(2+m)!$.
It turns out that the {\color{black}time derivative} of the angle $h_{\Moon}$ is well approximated by a constant frequency   defining a period of $18.6$ years. In other words, we consider the explicit {\color{black}time dependence} of the lunar potential {\color{black}as linear}. At this stage, it is recognisable and transparent that the Hamiltonian formed  on the perturbations,
\begin{align}\label{Eq:H}
	\mathcal{H} = \mathcal{H}_{\textbf{{\color{black}$J_{2}$}}}(G,H)  -\mathcal{R}_{\Sun}(G,H,g,h)-\mathcal{R}_{\Moon}(G,H,g,h,t), 
\end{align}
{\color{black}possesses $2$ DOF} and is periodically-time dependent (\ie a $2.5$-DOF problem).  The {\color{black}explicit time} dependence {\color{black} due only} to the node of the Moon\footnote{ 
We emphasise that the Hamiltonian depends on time just through the lunar contribution as we assumed that, over our timescale of interest, the rate of variation of the ascending node of the Sun is zero {\color{black}(see discussions in \cite{aCe17})}. 
} plays a fundamental role in shaping the dynamics. 
The well-known distinctive feature with the case of {\color{black}$2$ DOF} is that, \textit{a priori} (in absence of additional known first-integrals {\color{black}apart the energy function itself}), the tori cannot act as  practical barriers preventing transport in the phase space (for an $N$-DOF autonomous problem with $N \ge 3$, the codimension between  the $N$-dimensional tori and the dimension of the {\color{black}phase space} restricted to an {\color{black}energy surface} ($2N-1$) is at least $2$). The Delaunay variable $\ell$ being a cyclic variable, its conjugate variable $L=\sqrt{\mu a}$ is a constant of motion. Let us introduce normalised new actions $\tilde{x}=x / \sqrt{\mu a}$. The {\color{black} reduced} system is kept canonical as long as the new angles $\tilde{y}=\sqrt{\mu a} \cdot y$ are introduced and the physical-time multiplied by the same factor.  
It is clear that the new Hamiltonian has the same form as in Eq.\,(\ref{Eq:H}). The previous factor $\alpha_{J_{2}}$ absorbs now a contribution from $L$ and we get the new  $\alpha_{J_{2}}=J_{2}r_{\Earth}^{2}\mu^{4}/4L^{6}$. Factorizing the external perturbation by the greatest $\alpha_{\Moon}$, the Hamiltonian can be rewritten as
\begin{align}\label{Eq:Hnew}
	\mathcal{H}(\tilde{G},\tilde{H},\tilde{g},\tilde{h},\sqrt{\mu a}t) =
	\underbrace{\alpha_{J_{2}} f_{0}(\tilde{G},\tilde{H})}_{
	h_{0}(\tilde{G},\tilde{H})}
	 + \alpha_{\Moon}
	\underbrace{
	\Big(
	-\frac{\alpha_{\Sun}}{\alpha_{\Moon}} \tilde{\mathcal{R}}_{\Sun}(\tilde{G},\tilde{H},\tilde{g},\tilde{h})
	-
\tilde{\mathcal{R}}_{\Moon}(\tilde{G},\tilde{H},\tilde{g},\tilde{h},\sqrt{\mu a}t)
	\Big)}_{
	h_{1}(\tilde{G},\tilde{H},\tilde{g},\tilde{h},t)
	}
	.
\end{align}
The hierarchy $\alpha_{\Moon} \ll \alpha_{J_{2}}$ enables us to write $\alpha_{\Moon} = \varepsilon \alpha_{J_{2}}$, $\varepsilon \ll 1$, and Eq.\,(\ref{Eq:Hnew}) becomes
\begin{align}\label{Eq:Hfinal}
	\mathcal{H}(\tilde{G},\tilde{H},\tilde{g},\tilde{h},\sqrt{\mu a}t) =
	h_{0}(\tilde{G},\tilde{H}) + \varepsilon h_{1}(\tilde{G},\tilde{H},\tilde{g},\tilde{h},t).
\end{align} 
Clearly $\mathcal{H}$ shares the form of {\color{black}the} standard perturbed Hamiltonian system as announced in the introduction. 
The very useful information that we got from these manipulations is that the dimensionless perturbative parameter $\varepsilon$ is proportional to the secular invariant semi-major axis, 
\begin{align}
	\varepsilon(a) \equiv \frac{\alpha_{\Moon}}{\alpha_{J_{2}}} = \frac{2n^{2}_{\Moon}}{(1-e_{\Moon}^{2})^{3/2}} 
	\cdot
\frac{1}{J_{2}n^{2}} 
\Big(
\frac{a}{r}
\Big)^{2}.
\end{align}
({\color{black} The mean motions of the test particle and disturbing bodies are noted $n$ and $n_{\Moon}$, respectively.}) Note that this perturbing parameter is of the same nature as {\color{black} that} introduced by \cite{sBr01}, {\color{black}but we are treating herein the regime of the lunisolar {\it secular} (not {\it semi-secular}) resonances}. 
The Hamiltonian model based on the quadrupolar level is physically relevant up to a semi-major axis close to  $a_{\max} = 6 r_{\Earth}$ (beyond, octupolar refinements, $l=3$, are needed) corresponding to $\varepsilon(a_{\max})=0.8$. In the following, we will be interested in semi-major axes up to $a=29,600$ km, leading to $\epsilon=0.22$.
From our numerical investigations, we noted that  for $a=13,600$ km, the chaos is thin and confined to a few inclination-dependent-only resonances. These two constraints together define the subinterval $\mathcal{I}=[0.004,0.22] \subset \mathbb{R}_{+}$ of interest for $\varepsilon$. Adding quite `virtually' the point $\{0\}$ to this set,  $\varepsilon \in \Big(\{0\}\bigcup \mathcal{I}\Big)$, we obtain  when $\varepsilon=0$ an integrable dynamics with a linear flow on a torus. 
The actions are constant and determine the invariant tori. On these tori the dynamics consist of a rotation at constant speed characterised by the vector of constant frequencies (\textit{the unperturbed frequency vector}) $   \Omega(G,H)=(\varpi_{g},\varpi_{h})$ given by 
\begin{align}
\left\{
	\begin{aligned}
	\varpi_{g} & = \frac{1}{2} \kappa (5 \cos^{2}i-1) (1-e^{2})^{-2},\\
	\varpi_{h} & = - \kappa \cos i (1-e^{2})^{-2},
	\end{aligned}
\right.
\end{align} 
where $\kappa = \frac{3}{2} J_{2} n r_{\Earth}^{2}/a^{2} \in \mathbb{R}$ \citep{wKa66}. 
{\color{black} 

Let us be more precise regarding the definition of the interval $\mathcal{I}$ and the energy function considered. An important factor leading to the $2.5$ DOF model lies in the omission of the \textit{tesseral} contributions in the Hamiltonian (\ref{Eq:H}).  
Tesseral resonances occur when the commensurability 
$\dot{M}/\dot{\theta} \sim q/p$ takes place. 
Given the upper and lower bounds of $\mathcal{I}$, the  \textit{main} tesseral resonances affecting the motion
are given by the set of commensurabilities $\mathcal{T}=\{
2:1, 3:1, 4:1\}$ \citep[see \eg][]{tEl96,aCe15minor}.   
Near a tesseral resonance, the semi-major axis is no longer 
secularly invariant and, in fact, might experience (confined) chaotic variations near their corresponding resonant action value $L_{\star}^{p:q}$. 
Putting all of these together, a more precise definition of our interval of interest for the perturbing parameter $\varepsilon$ is  instead 
$\mathcal{I}^{'}=\mathcal{I} \, \setminus \, \mathcal{I}_{\mathcal{T}}$ where
$\mathcal{I}_{\mathcal{T}} = \bigcup_{p:q \,\in \, \mathcal{T}} [\varepsilon(L_{\star}^{p:q} - \delta_{p:q},\varepsilon(L_{\star}^{p:q}) + \delta_{p:q}]$ where $\delta_{p:q}$ characterise the strength (width) of the resonance. The eventual local coupling of the tesseral contributions  and the lunisolar resonances on the evolutions of the actions $(G,H)$, for perturbing values precisely within the tesseral resonant domain (\ie for $\varepsilon \in [\varepsilon(L_{\star}^{p:q} - \delta_{p:q},\varepsilon(L_{\star}^{p:q}) + \delta_{p:q}]$), is not discussed in this contribution, but is the object of a further study (accordingly, tesseral contributions are disregarded in the energy function). 
Using estimations obtained in former works \citep{tEl96,aCe15minor}, let us stress that the ratio of the measures 
$\textrm{m}(\mathcal{I'})/\textrm{m}(\mathcal{I}) \sim 0.99$  is very close to $1$. 
Moreover, accurate low-altitude or very local studies could require additional refinements of the energy function \citep[\eg near the critical inclination, see][]{mLa18}. 
Although our model is subject to these possible limitations, we should highlight that the considered dynamics are accurate enough to describe general MEO orbits, and from a mathematical point of view provide a simple testbed to investigate transport theories and capture the big picture.}

We abused the vocabulary and {\color{black}treat} the eccentricity $e$ and inclination $i$ as `actions', instead of {\color{black}using} the veritable actions variables $(G,H)$.
This is rather to stick to classical notations \citep{wKa66}. Nevertheless, these variables are functionally independent and {\color{black}the `true' actions} easily obtained as $e^{2}=1-(G/L)^{2}$ and $\cos i=H/G$. 
Let us precise that, when dealing with the autonomous Hamiltonian by introducing an {\color{black}extra} conjugated variables $(\Gamma,\gamma) \in \mathbb{R}\times\mathbb{T}$ for $\varepsilon \neq 0$, one couple of actions $x=(G,H)$ characterise an invariant torus of $\mathbb{T}^{3}$ since $\Gamma$ does not enter into the equations of motion. 
In other words, we can consider the orbits in the reduced phase space defined  by  $\mathcal{X}=\big\{ (G,H,g,h,\gamma), x=(G,H) \in D \subset \mathbb{R}^{2}, y=(g,h,\gamma) \in \mathbb{T}^{3}\big\}$. In Section~\ref{sec:FLIsurvey}, we will offer views of the dynamics in action-action sections, meaning that within the space $\mathcal{X}$, we fix the angles to a specific vector to obtain the section
$S=\big\{ (G,H) \in D \subset \mathbb{R}^{2} \, \vert \, (g,h,\gamma)= v_{\star},\, v_{\star} \in \mathbb{T}^{3}\big\}.$ Let us now discuss fundamental phenomenon for $\varepsilon \neq 0$.
 
%------------------------------------------------------------
\subsection{Secular Lunisolar Resonances}
%------------------------------------------------------------
A determinant feature in the long-term properties of nearly-integrable systems of the form 
$h(x,y) = h_{0}(x) + \epsilon h_{1}(x,y)$ is the presence of resonances\footnote{In the `multiscale analysis' community, resonances are sometimes named `\textit{slow hidden variables}', see \eg \cite{gAr09,aAb12}. This semantic is pretty accurate as this is precisely what resonances are: resonances form `new slow variables' solely  under specific combinations of the fast variables. Having this in mind, it is clear that in the presence of resonances the \textit{direct averaging} may be crude (`naive' averaging) and conducts to a wrong dynamics. 
} \citep{pLo02}. They arrive when a vector $k \in \mathbb{Z}_{\star}^{n}$
satisfy with the (unperturbed) frequency vector  a commensurability condition over the {\color{black}rationals}. 
The resonant condition reads  
$k \cdot \Omega(x)  = 0$. For a fixed vector $k \in \mathbb{Z}_{\star}^{n} $, the sets (potentially empty) of the actions $x$ such that $k \cdot \Omega(x) = 0$ form the \textit{resonant manifolds}.  The resonance under consideration is then characterised by an index, \textit{the resonan{\color{black}ce} order}, usually though the $\ell_{1}$-norm of $k$,
$ \norm{k}_{1}= \sum_{i}\vert k_{i}\vert$. 
Under the quadrupolar assumption, the system (\ref{Eq:Hfinal}) is prone to resonate
with a maximal order of $6$. Let us consider the frequency vector $\Omega(x)=(\varpi_{g}(x),\varpi_{h}(x),\varpi_{\Moon})$, then, as already recognised by \cite{tEl96}, the resonant conditions read as
\begin{align}\label{Eq:ResCond}
	k_{1} \varpi_{g}(x) + k_{2} \varpi_{h}(x) + k_{3} \varpi_{\Moon} = 0, \, k_{1} \in \{-2,0,2\}, \, k_{2} \in  \llbracket 0,2\rrbracket, \, k_{3} \in  \llbracket-2,2\rrbracket, \, k \neq 0.
\end{align}
These algebraic equations admit non-trivial solutions that define the lunisolar resonant manifolds.  The resonant manifolds are mirrored with respect to the resonance $(0,k,0) \cdot \Omega(x)=k\varpi_{h}(x)=0$. (However, as we will clearly illustrate it, the symmetry of the resonant manifolds does not imply a mirroring of the geography of the KAM tori and hyperbolic structures.)  
%Just this once, properties of the phase-space will be depicted in the eccentricity and inclination phase-space. 
%These 'actions' variables 
%are not actions in the proper sense, however they are directly related( and functionally independent) to the proper Delaunay actions by $\cos i=H/G$ ($0 \le i < \pi$) and $e^{2}=1-(G/L)^{2}$.
%The use of these 'actions' variables facilitate the physical interpretations of the results.
%That said, we sometimes abuse of the vocabulary and refer to these two observables as actions, without the extra care of using the quotation marks.  
In \cite{jDa16}, the {\color{black}extent of the resonant zones have been estimated (in a subdomain of the prograde $0< i \le \pi/2$ domain) 
by reducing the Hamiltonian to the \textit{first fundamental model} of resonance, a pendulum. 
This procedure involved the introduction of  resonant coordinates through canonical transformations $\frak T_{k} \in \textrm{SL}(3,\mathbb{Q})$} {\color{black}leading to an intuitive physical interpretation of  Chirikov's overlap as a driver of chaos.} 
{\color{black}However}, since the work of \cite{aCe16-siam}, it has been observed that such a reduction {\color{black}does not always capture the features of the dynamics. In order to get a more refined and precise view of the extent of chaos, a superior way is instead to look at the destruction of KAM curves, \eg using fast dynamical indicators.} 

%Nevertheless, the interactions of nearby resonances constitute most probably the driver of the apparition of chaos and this enterprise gave us a physical intuition behind the transition from oder to chaos when $\varepsilon$ sweeps its domain of defintion.   Let us now build further on this transition from order to chaos.

%----------------------------------------------------
\section{Phase-space views}\label{sec:FLIsurvey}
%----------------------------------------------------
We revisit and complement the transition order/chaos in terrestrial orbits by scanning the dynamics under the rays {\color{black} of a first order variational chaos indicator}.  The motivation is twofold: 
\begin{enumerate}
	\item the resolutions used in former studies are generally sufficient to detect and isolate chaotic components; yet, they are too coarse to detect the eventual presence of structures immersed within them. In addition, the dissection of the dynamics with a fine resolution makes possible the extraction of the chaotic skeleton with  surgical precision\footnote{In a somehow different but connected context, \cite{rAr18} have shown that fine discretisations are also needed for optimisers to operate properly.}. This property {\color{black} will be used} to study  transport 
properties (\textit{confer} Section~\ref{Sec:DriftTransport}).
	\item \cite{iGk16} claimed that '\textit{the retrograde orbits are not intrinsically more stable then their prograde counterparts}'. This 
diagnosis was established by scanning the region with an \textit{averaged} FLI (over some angles) focusing on low eccentricity (up to $e=0.1$). We feel necessary to investigate further this {\color{black} assertion} beyond $e=0.1$ without the {\color{black}averaged} indicator {\color{black}(}{\color{black}which} naturally tends to smooth and absorb the details{\color{black})}. 
\end{enumerate}	 
To overcome and constrain these two symptoms, 
we first briefly recall how we efficiently deal with the equations of motions,  {\color{black}after which}  we  present and discuss  our highly-resolved phase-space views on a macroscale. 

%----------------------------------------------------
\subsection{Numerical treatment of the equations of motions}\label{SubSec:EDOs}
%----------------------------------------------------
Ordinary and partial differential equations with disparate scales (spatial, temporal)  are numerically challenging. The difficulty arrises from the fact that the inhomogeneities in the scale constrain parameters of the numerical methods employed (say, \eg the size of the timestep, the discretisation of the mesh)  to be small and highly resolved. 
In the present case, we deal with (highly) oscillatory ODEs. They are omnipresent in the context of Newton's equations\footnote{{\color{black}They arrive also in Molecular Dynamics (see \cite{mAl04} and \cite{gAr98} for introductory papers and issues).}} and are ubiquitous in the context of Celestial Mechanics.
To circumvent the problem, \textit{effective models} and \textit{model reductions} techniques are often employed. The core idea is to substitute to the original dynamics a more amenable, numerical and/or analytical, dynamical system \citep{dGi04,aLe06,cHa07}.  
One method of choice to design effective dynamics relies on the Lagrangian \textit{averaging principle} \citep{eGr65,iMi67,eGh07,gPa08}, which has a long-lasting tradition in Celestial Mechanics.  
This principle is usually used when the components of the equations themselves allow to recognise explicitly the time scales. When it is so, the fast dynamics is \textit{integrated} into the slow variables to design an \textit{averaged} approximation.  In this setup, the new slow constituents somewhat incorporate the informations of the fast-dynamics and serve as a new input for the investigations.  
To deal efficiently with our problem at hand, we adopt here our own secular \texttt{MILAN} model, as presented by \cite{iGk16}. The \texttt{MILAN} model is based on the vectorial Milankovitch element and admits a minimal force model (consisting of the averaged $J_{2}$ contribution, to which is added the secular quadrupolar third-bodies perturbations). The \texttt{MILAN} formulation bears also net advantages compared to the numerical treatment of the Hamiltonian equations in forms of those given in Section~\ref{subset:model}. First, the formulation is free of singularity and, secondly, the averaging  is done in a closed form in the eccentricity.
The external third-bodies potentials are both averaged over the fast variables of the problem, \ie over 
the mean-anomaly of the test particle \textit{and} the mean anomalies of the third bodies. This `doubly' averaged model {\color{black} allows} the propagation of a test-particle over $10^{6}$ to $10^{7}$ orbital periods in a few seconds only. Such a performance is essential in investigating properties of the phase space for range of parameters.  
When invoking effective models, we always face the question of the relevance of the reduced model (how sound are the qualitative or quantitative informations derived from it). \cite{iGk16} established the testimony of this doubly-averaged model against a singly-averaged approach. By simulating the two dynamics on different domains of the action-action, action-angle and angle-angle spaces\footnote{In \cite{iGk16}, we presented only sections in the angle-angle space but we have evidences of the agreement on complementary sections also for a range of different semi-major {\color{black}axes.}}, we showed that dynamical features of interests were reproduced and in perfect agreement. Even if the simulation of the full dynamics (\ie the original, non-reduced and 'exact' dynamics) on such domains is still missing in the literature, recent reassuring numerical agreements have been presented by \cite{rAr18}. Namely, they presented nice agreements between their in-house doubly average model and the original non-averaged dynamics. All these together allow us to be confident enough on the numerical  results presented hereafter.

%----------------------------------------------------
\subsection{Highly resolved phase-space views}\label{subsec:PSV}
\begin{figure}
\centering
\includegraphics[width=0.99\textwidth]{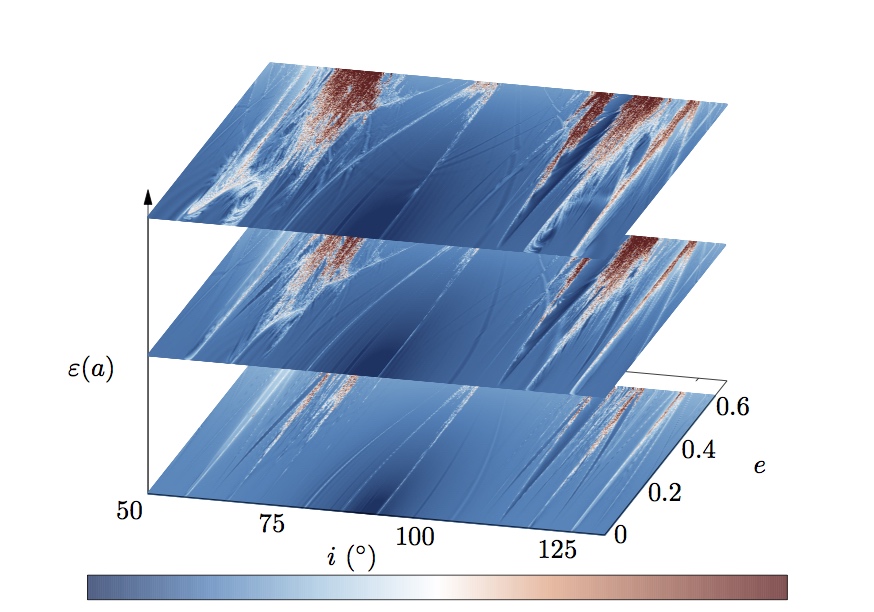}
\caption{\label{Fig:FenceProgradeFLIs-part1} A highly-resolved fence view of the stability of the prograde \textit{and} retrograde regions obtained under the FLIs.  The three slides depict the stability for a particular value of the {\color{black} non-linearity} parameter $\varepsilon(a)$ which depends on the secularly invariant semi-major axis. 
KAM tori correspond to {\it white} to {\it light-red} color, stable resonant orbits appear in {\it blue} while {\it red} colors correspond to chaotic orbits. 
The values $\varepsilon(a)$
correspond to the three semi-major axis $\{18.6,22.6,24.6\} \times 10^{3}$ km (the z-scale is only symbolic, in particular the scale is not linear).
This unscrewed view presents in a global, original and concise way the transition from order to chaos. See text for comments.
}
\end{figure}

\begin{figure}
\centering
\includegraphics[width=0.99\textwidth]{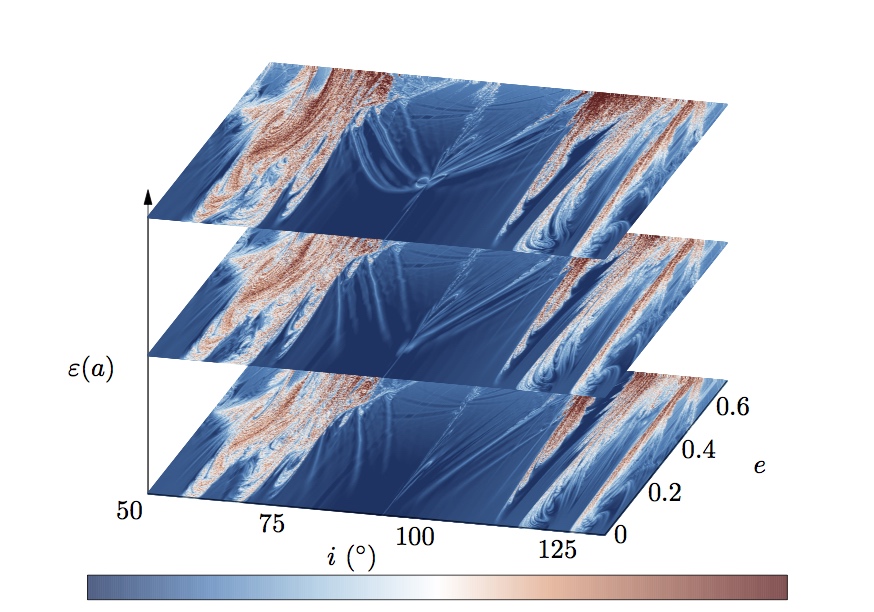}
\caption{\label{Fig:FenceProgradeFLIs-part2} The same as in Fig.\,\ref{Fig:FenceProgradeFLIs-part1}
apart that  
the values $\varepsilon(a)$
correspond to the three semi-major axis $\{27.6,28.6,29.6\} \times 10^{3}$ km.
}
\end{figure}

%----------------------------------------------------
 {\color{black}We use the Fast Lyapunov Indicator (FLI)}, a first-order variational indicator initially introduced by {\color{black}\cite{cFr97}}, to discriminate orbit stability. This scalpel has been used extensively over the past decade across 
different dynamical problems, ranging from Symplectic Maps studies to Dynamical Astronomy, including Astrodynamical practical problems \citep{cFr00,nTo15,nGu17,aRo17}. The work of C.\,Froeschl\'e, M.\,Guzzo and E.\,Lega over the last decade provides a good overview of its possibilities and range of applications. When the dynamics under consideration is  written in first order and autonomous form as $\dot{x}=f(x)$, $x \in \mathbb{R}^{n}$, the FLI is simply derived from the variational system  in  $\mathbb{R}^{2n}$,
\begin{align}
\left\{
 \begin{aligned}
 	&\dot{x} = f(x), \\
	&\dot{w} = \partial_{x}f(x)(w),
 \end{aligned}
 \right.
 \end{align}
 as
 \begin{align}
	\textrm{FLI}(x_{0},w_{0},\tau_{\textrm{run}}) = \sup_{0 \le t \le \tau_{\textrm{run}}} \log \norm{w(t)}.
\end{align}
Contrarily to Lyapunov exponents, the FLIs (computed at some time $\tau_{\textrm{run}}$ for a specific set of initial conditions  $x_{0},w_{0}$) keep trace of the resonant nature of the orbits, while taking approximately the same value 
$\textrm{FLI} \sim \log(\tau_{\textrm{run}})$ on KAM tori \citep{cFr00,mGu02}. 
By computing the FLIs on a discretised  specific $2d$-section  of ICs (\eg related to the action-action, action-angle or angle-angle planes) on a domain $D$, we can reveal the geography of the survival KAM tori and their complement hyperbolic set. The information given by the FLIs (`intensity') is then color-coded to obtain a \textit{map of stability}. Note that sometimes, in order to get a sharper visualisation, the FLIs {\color{black}that} initially take variation in $\mathcal{J}\subset \mathbb{R}_{+}$ are restricted to a subinterval $\mathcal{K}$ of $\mathcal{J}$ (see \eg \cite{eLe03,riPa15}). This rescaling is achieved by fixing the two following thresholds. 
The notion of chaoticity is based on the exponential evolution of the norm between two nearby orbits. Therefore, to reveal anomalies with respect to the linear trend (log-scale of an exponential growth), the criteria 
\begin{align}\label{Eq:FLIChaos}
\textrm{FLI}(\tau_{\textrm{run}}) \ge \log(\tau_{\textrm{run}}^{\alpha}) = \alpha   \log(\tau_{\textrm{run}}),\, \alpha >1,
\end{align}
can be used to derive a lower threshold for chaotic orbits (\ie all FLIs larger than $\alpha   \log(\tau_{\textrm{run}})$ are assigned to $\alpha   \log(\tau_{\textrm{run}})$). Symmetrically, we obtain an  upper threshold to judge regularity with the criteria 
\begin{align}
\textrm{FLI}(\tau_{\textrm{run}}) \le \log(\tau_{\textrm{run}}/{\beta}),\, \beta >1.
\end{align} 
(And again, all the FLIs smaller than $ \log(\tau_{\textrm{run}}/{\beta})$ are assigned to $\log(\tau_{\textrm{run}}/{\beta})$.)
The Figs.\,\ref{Fig:FenceProgradeFLIs-part1} and \ref{Fig:FenceProgradeFLIs-part2} resume in many ways the transition from order to chaos in the prograde \textit{and} retrograde region of terrestrial orbits. This original unscrewed fence views of the action-action phase space are very illuminative (and pedagogical) 
to visualise, with respect to the non-linear parameter $\varepsilon(a)$,  the 
proliferation of chaos. 
Each map represents the result of  $1,000 \times 500$ ICs 
propagated over a {\color{black}long timescale.}  Different simulation times $\tau_{\textrm{run}}$ have been used according to the perturbing parameter (the stronger is the perturbation, the shorter is the time required to get a sharp contrast of the dynamics). For the smallest perturbing parameter, $a=18,600$ km, $\tau_{\textrm{run}}$ represents $30$ lunar nodes, while for 
$a=29,600$ km $16$ lunar nodes are sufficient to get a sharp contrast (most probably those propagation times could be slightly shortened). It represents about $7 \times 10^{5}$ and $1.8 \times 10^{5}$ test {\color{black}particle revolutions}. The `actions' have been uniformly distributed along the rectangle 
$[50^{\circ},130^{\circ}] \times [0,e_{\max}]$, with $e_{\max}$
determined by the apogee-altitude condition 
$e_{\max}=1-(r_{\Earth}+\delta)/a$,  $\delta = 120$ km. In all our maps, we have  set the initial angles $y_{0} \in \mathbb{T}^{3}$ to zero.  Anticipating a bit the next section, we are here interested in the  dynamical mechanisms leading to transport; in particular we have not discriminated  \textit{collisional orbits} as we did in earlier work.   As it has already been discussed several times and pointed out in several contributions \citep{jDa16,iGk16,aCe16}, the inclination dependent-only resonances
 widen and develop chaos when $\varepsilon$ is increasing, letting less and less room for invariant KAM tori. 
Eventually for $a=29,600$, there is a macroscopic chaotic component. At this macroscale, we even have the feeling of a \textit{chaotic path-connected space} (\ie for every two points in the hyperbolic set, there exists a hyperbolic path connecting them). This property is not exactly true as isolated chaotic islands do exist. Nevertheless, the volume of such isolated chaotic sea  is rather small. 
Let us precise that, given the fact that we used different $\tau_{\textrm{run}}$, the color palette has a symbolic meaning only. Also, in the same way, the $z$-scale which sets the different levels of the perturbing parameter $\varepsilon(a)$ is not a linear scale, and again, has only a schematic pictorial purpose.

\begin{figure}
\centering
\includegraphics[width=0.8\textwidth]{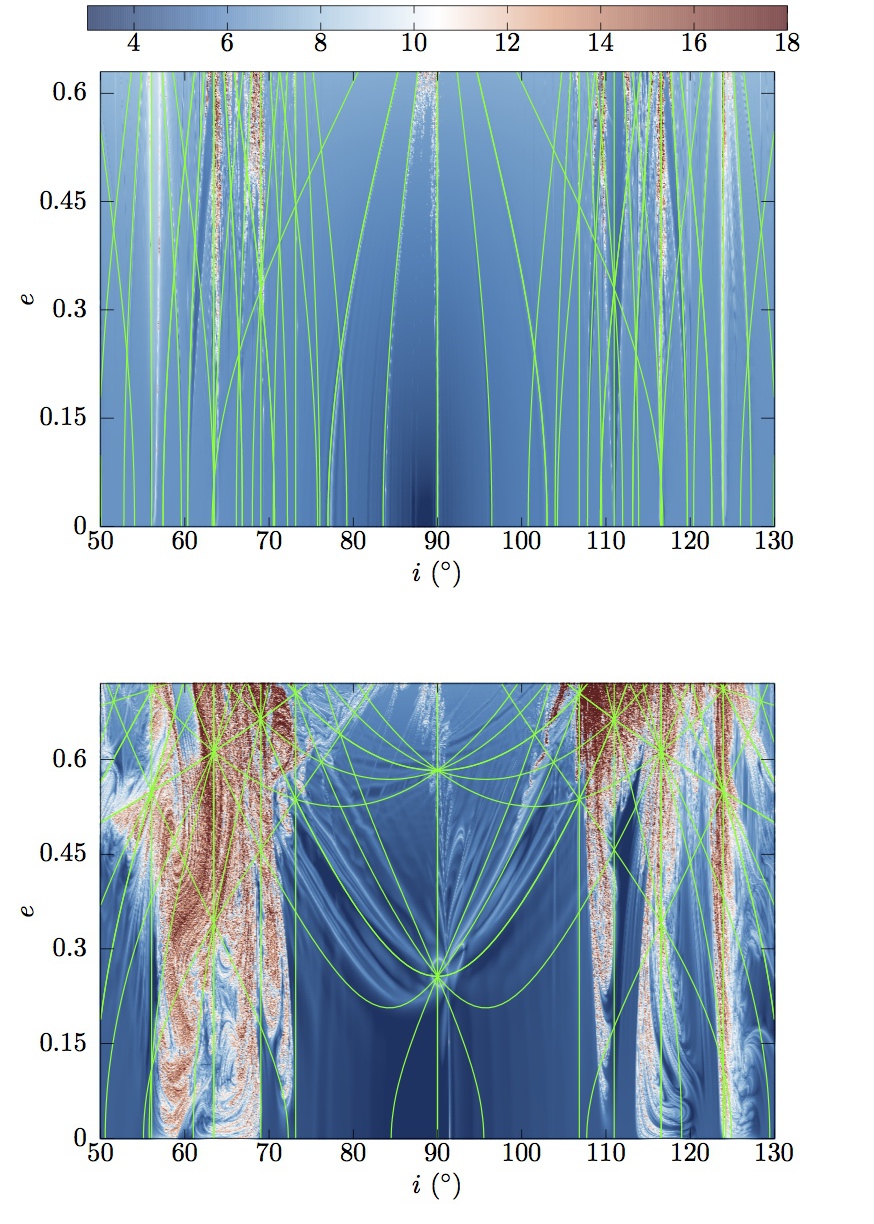}
\caption{\label{Fig:PRwManifolds} 
Detailed views of the prograde \textit{and} retrograde regions for the two-extreme values of the parameter $\varepsilon$ considered in this work ({\color{black}top panel $a=18,600$ km, bottom panel $a=29,600$ km}). 
The resonant manifolds defined by Eq.\,(\ref{Eq:ResCond}) are superimposed on the FLIs. Despite the symmetry of the resonant manifolds, the chaos of the prograde region is not mirrored in the retrograde region.}
\end{figure}

\begin{figure}
\centering
\includegraphics[width=0.8\textwidth]{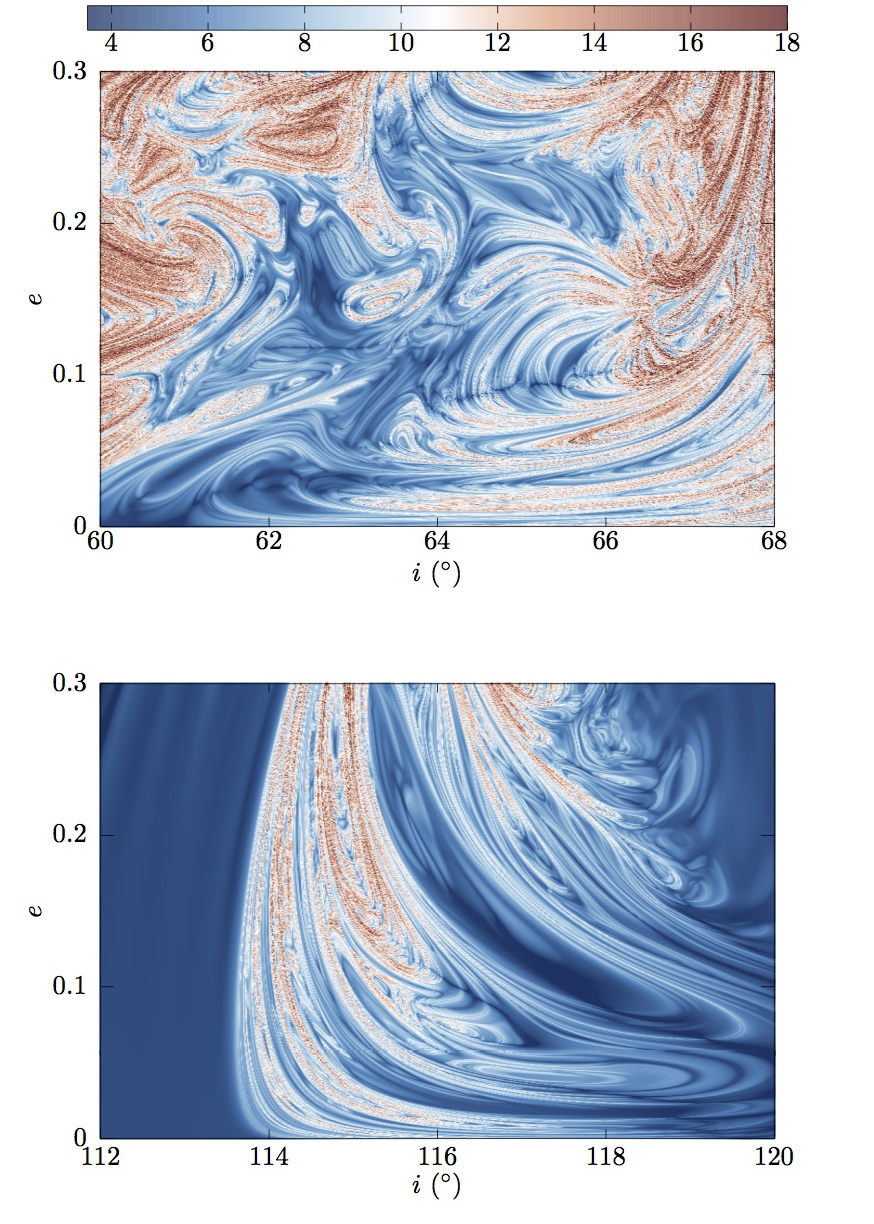}\\
\caption{\label{Fig:MicroProRetro}
Two detailed views of the phase space under the FLI analysis. 
$701 \times 701$ orbits have been propagated revealing the 
existence of very thin structures and KAM tori filling the chaotic regions. On this domain, the retrograde region contains $4$ time less orbits satisfying the condition $\textrm{FLI}(\tau_{\textrm{run}}) \ge \alpha \log(\textrm{run})$, $\alpha=1.1$. 
}
\end{figure}

For the two extreme perturbing parameters considered in this work, $\varepsilon(a)$ with $a=18,600$ or $29,600$ km, we have superimposed for the newcomers the resonant manifolds obtained under the quadrupolar assumption (confer Eq.\,(\ref{Eq:ResCond})).  It is interesting to notice  that,
despite the symmetry of the resonant manifolds along the $(0,k,0)$ resonance, the chaos is not mirrored at all in the retrograde region (see Fig.\,\ref{Fig:PRwManifolds}). 
{\color{black}The coefficients of each harmonic, excepting the critical inclination, are dependent on the cosine of the inclination and hence the resonant topologies for prograde and retrograde orbits are necessarily different.}
A further striking illustration of this fact, on a microscale, is exemplified in Fig.\,\ref{Fig:MicroProRetro}. Such  fine resolutions allow to reveal incredible structures and details of the phase space. These two simulations clearly show us, at least for this realisation of angles, that the retrograde region is more stable than its prograde counterpart. Applying the criteria given by Eq.\,(\ref{Eq:FLIChaos}) with $\alpha=1.1$, we found on that domain that the volume of chaotic orbits is $4$ times larger than in its retrograde counterpart.  
We further quantified this question by applying various criteria on our former simulations. 
Tab.\,\ref{Tab:ProVSRetro} summarises our results by giving the volume of chaotic orbits in the prograde \textit{versus} the retrograde region,    
for slightly different values of $\alpha$ on a macroscale.  From our survey (which should be extended for completeness), the numerics tend to show that, for
small to moderate values of the perturbing values of $\varepsilon$, the 
volume of chaotic orbits is roughly the same. However, for larger values of $\varepsilon$, the prograde region is more chaotic than its counterpart, the difference being now of several percent. 
(But again, we are aware of the dependence of our result against the choice of $y_{0}$.
Further numerical investigations could constrain even more the result.)
At a smaller scale, as already recognisable  in Fig.\,\ref{Fig:MicroProRetro}, the discrepancies may be largely more significant.  
It would be interesting to support or invalidate this phenomenology by characterising i) the widths of the resonances of the retrograde domain and ii) by exploring their numerical widths as a function of the angles. Such an enterprise is {\color{black}yet} to be performed.

\begin{figure}
\centering   
   \includegraphics[width=1\textwidth]{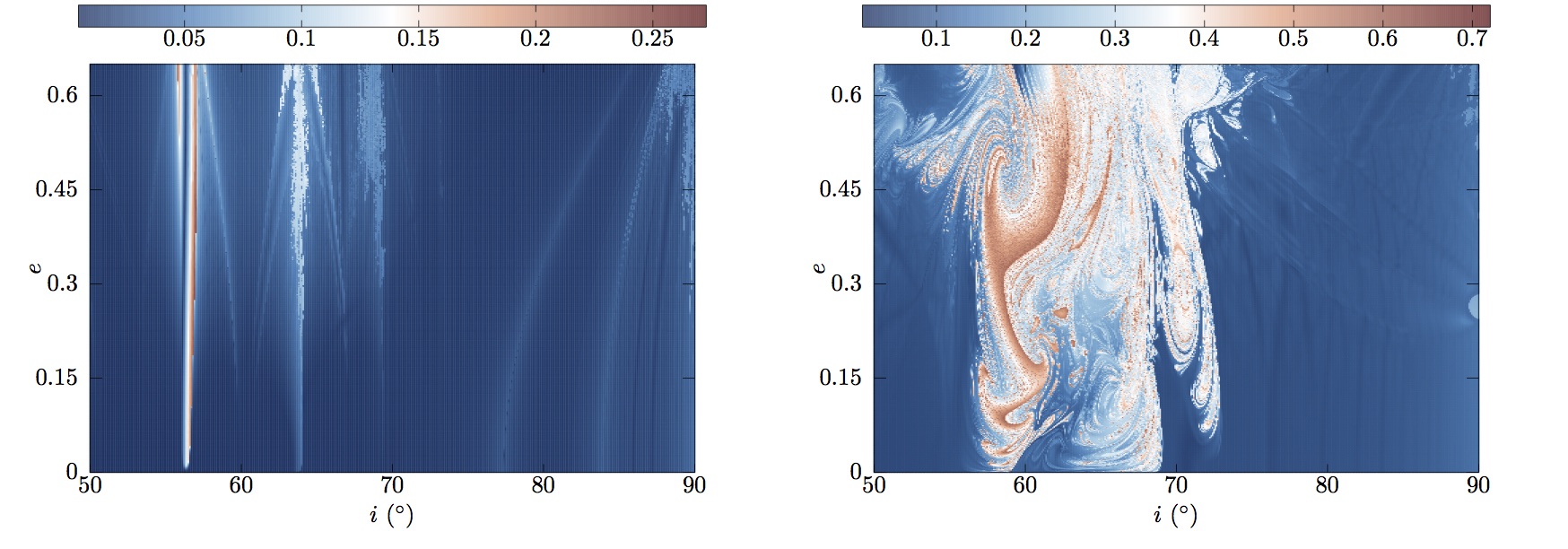}
 \caption{\label{fig:Diameters} 
Estimation of the diameters for the two perturbing parameters $\varepsilon(a)$ with $a=18,600$ km and $a=29,600$ km.
} 
\end{figure}

\begin{figure}
\centering
   \includegraphics[width=1\textwidth]{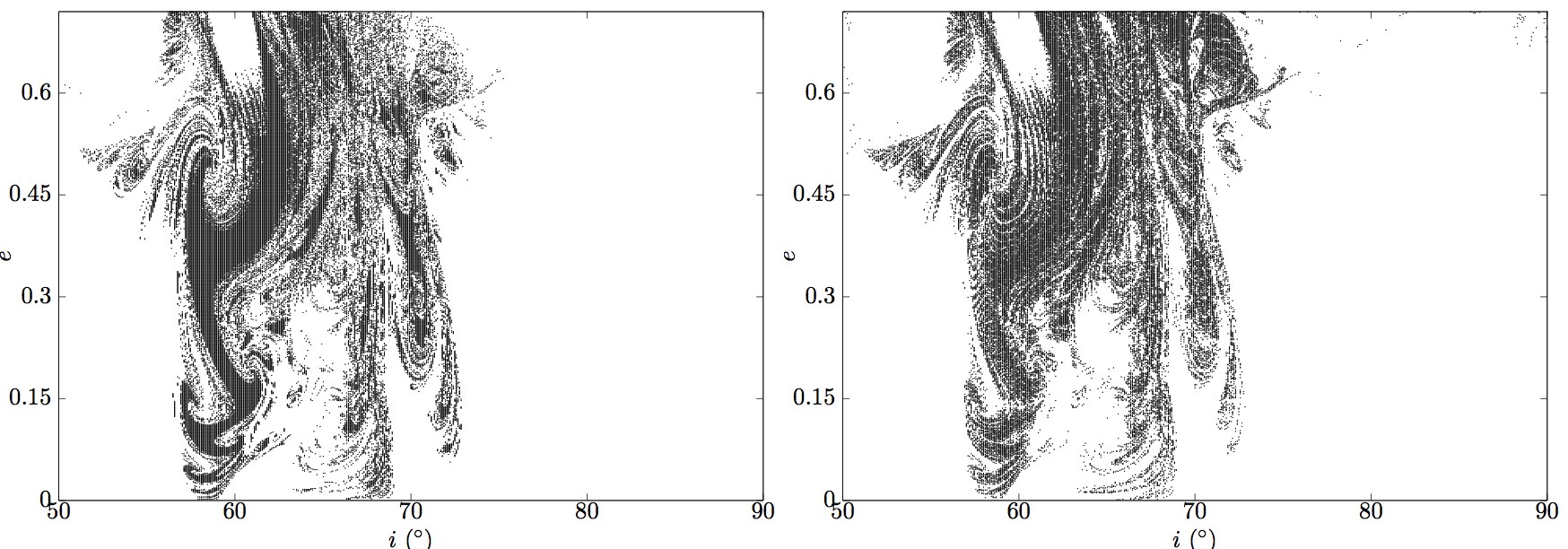} 
  \caption{\label{Fig:FLIvsD} 
Extraction of the ICs satisfying $\textrm{D}(x,\tau_{\textrm{run}}) \ge 0.35$ (left hand-side) and the chaotic ICs satisfying 
$\textrm{FLI}(x,\tau_{\textrm{run}}) \ge 1.2 \log (\tau_{\textrm{run}})$.
} 	
\end{figure}

\begin{figure}
\centering
\includegraphics[width=0.9\textwidth]{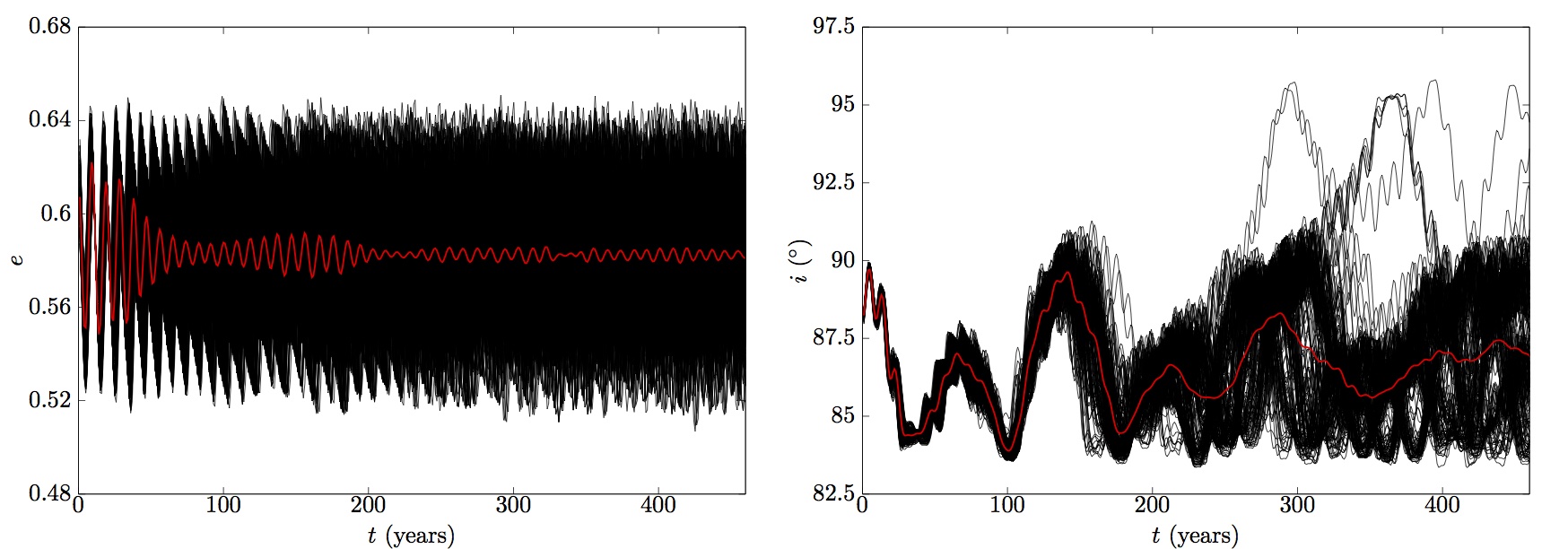}
\caption{\label{fig:EnsembleDiameter} Ensemble integration of a cluster of $k=200$ orbits in a neighborhood $\mathcal{V}(x_{\star})$ of  $x_{\star}\!=\!(0.616,88^{\circ})$. The ensemble averaged orbit of the considered observables are shown in {\it red}. 
The eccentricities do not experience a net sensitive dependence to the ICs, contrarily to the inclinations. From this example, it can be easily {\color{black}inferred} that a diameter-measure based solely on the eccentricity (or equivalently on $G$) would fail to capture properties of the dynamics.}
\end{figure}

\begin{table}
\centering
\setlength{\arrayrulewidth}{1.2pt}
\setlength{\arrayrulewidth}{1.2pt}
\begin{tabular}{c|cccc}
\hline
\multirow{2}{*}{Domain $D$}	&\multirow{2}{*}{Size of $a$ fixing $\varepsilon(a)$ [km]} & \multirow{2}{*}{Threshold $\alpha$} & \multicolumn{2}{l}{Volume of chaotic orbits} \\ 
	&							  & &  $\mathcal{V}_{\alpha}^{+}$ & $\mathcal{V}_{\alpha}^{-}$ \\
 \hline
 \hline
\multirow{3}{*}{$[0,0.65]\times[0:\pi/2]$}  & \multirow{3}{*}{$18,600$} & 		$1.1$ 	& 0.009 & 0.009  \\ 
  &  & 	$1.25$ 	& 0.005 & 0.005  \\ 
  &  & 	$1.3$ 	& 0.004 & 0.005  \\ 
\hline
\multirow{3}{*}{$[0,0.74]\times[0:\pi/2]$}  & \multirow{3}{*}{$24,600$} & $1.1$ 		& 0.05  & 0.05 \\ 
  &  & $1.25$ 	& 0.038& 0.04  \\ 
  &  & $1.3$ 	& 0.035 & 0.038 \\ 
\hline
\multirow{3}{*}{$[0,0.76]\times[0:\pi/2]$}  & \multirow{3}{*}{$27,600$} & 	   $1.1$  	& 0.18 & 0.12  \\ 
  &  & $1.25$  	& 0.12 & 0.09 \\ 
  &  & $1.3$    	& 0.1 & 0.08  \\ 
\hline
\multirow{3}{*}{$[0,0.7{\color{black}8}]\times[0:\pi/2]$}  & \multirow{3}{*}{$29,600$} & $1.1$ 		& 0.22 & 0.14 \\ 
  &  & $1.25$ 	& 0.16 & 0.09 \\ 
  &  & $1.3$ 	& 0.14 & 0.08 \\ 
 \hline
 \hline 
\end{tabular}
\caption{\label{Tab:ProVSRetro}
{\color{black}The} estimation of the volume of chaotic orbits in the prograde \textit{and} retrograde regions, for various perturbing parameters \textit{and} on various domains.  The domain $D$ refers to the definition of the domain in the prograde region, in the  eccentricity-inclination action phase space. This  domain is then mirrored in its retrograde counterpart to serve as a new domain to determine the volume of chaotic orbits in the retrograde region, $\mathcal{V}_{\alpha}^{-}$. {\color{black}The Eq.\,(\ref{Eq:FLIChaos}) is used as a discrimination criteria.}
All results have been established with a fine mesh (all domains have been uniformly discretised with a grid consisting of  \textit{at least} $500 \times 500$ initial conditions). 
The prograde region appears to be slightly more chaotic than the retrograde counterpart on a macroscale the more we increase the perturbing parameter. Significant differences may also exist at smaller scales.
}
\end{table}

\begin{figure}
\includegraphics[width=1\textwidth]{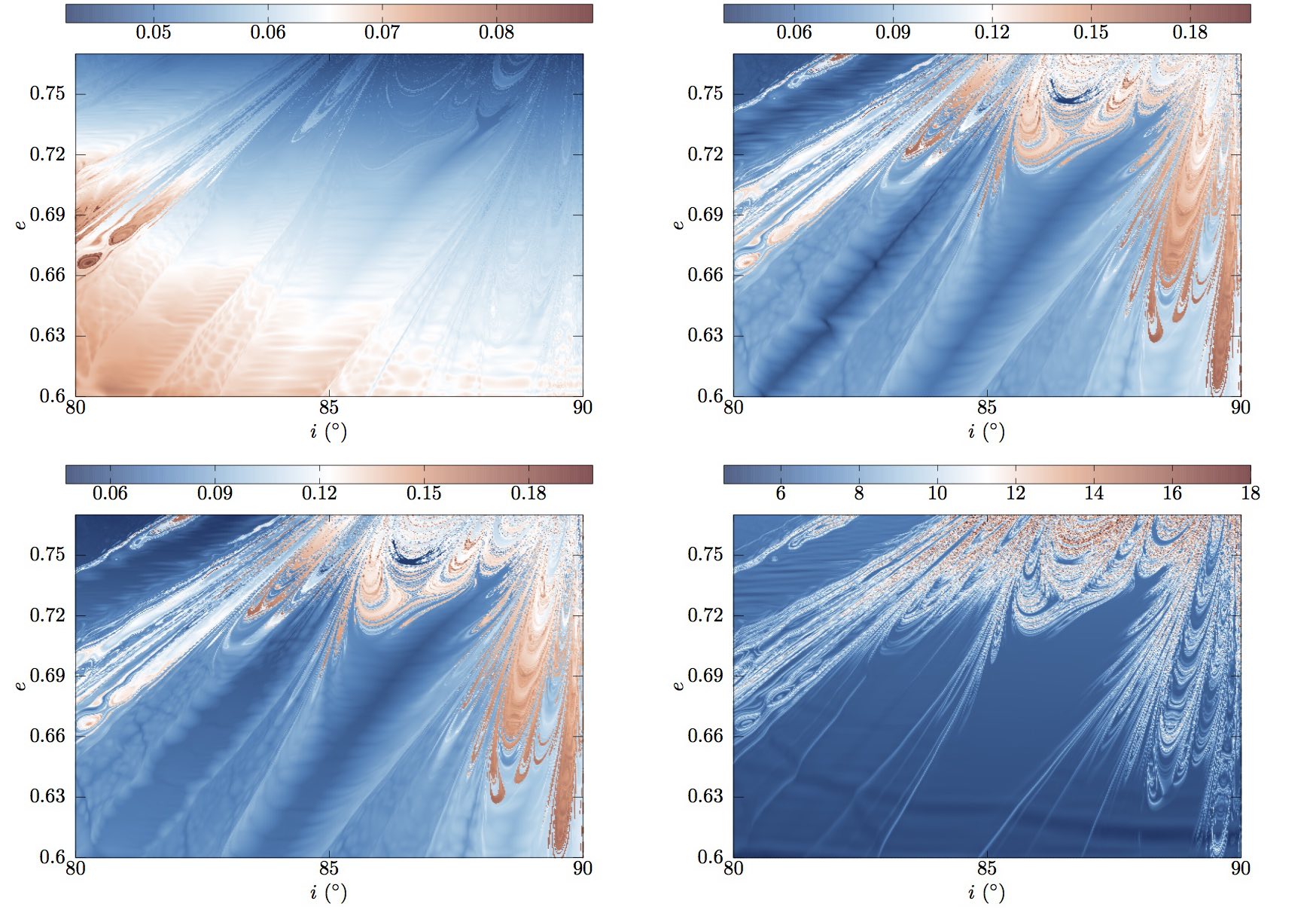} 
\caption{\label{fig:D1dvsD2d} The figures enable us to quantify how a one-dimensional diameter coefficient (here based on the action $G$, top left) may be inappropriate in some cases in capturing dynamical properties. \textit{A contrario}, the two-dimensional diameter coefficient (bottom left) based on both actions, $G$ and $H$, captures the subtleties of the dynamics  and reconcile the results with the FLIs analysis.} 	
\end{figure}

%---------------------------------------------------
\section{Drift and visualisation of transport}\label{Sec:DriftTransport}
%---------------------------------------------------
The computation of the FLIs provided a quantification of the degree of hyperbolicity and a discrimination of orbit stability. From a `practical' perspective, one might be more interested in drift estimation and visualisation of  transport  to {\color{black}quantify} changes of the  unperturbed first integrals.  
This section is devoted to this task by investigating asymptotic properties of initial conditions close or immersed in hyperbolic structures. We base our approaches on individual propagations and on spatial ensemble averages.

%---------------------------------------------------
\subsection{Drift estimation}
%---------------------------------------------------
There exist a tension between the local degree of hyperbolicity and the eventual large transport. 
In fact, the astronomical concept of \textit{stable chaos} 
teaches us that  positivity of a Lyapunov exponent does not necessarily implies large excursion 
in the phase space \citep{aMi92}. Large excursions in the phase space can be the signature of  
transport along the level curves of an integrable system. Nekoroshev's long-time stability 
theorem does not exclude the existence of chaotic variation.  Finally, beyond a critical value, 
Chirikov's overlap criterion of resonances 
give rise to large connected chaotic domains, allowing possibly macroscopic transport 
\citep{bCh79}. 

The problem of  chaotic transport (sometimes referred as \textit{chaotic diffusion}) in nearly-integrable Hamiltonian systems and Dynamical Maps still occupy efforts of various dynamicists 
(see \eg \cite{sLa16,nGu17,pCi18}). 
Given an orbit  computed up to a final time $\tau_{\textrm{run}}$, $\gamma(t)=\big\{\big(x(t),y(t)
\big)\big\}_{0 \le t\le \tau_{\textrm{run}}}$, we use the diameter along the action-variables  to 
measure the drift of the unperturbed first-integrals. More precisely, given an action-like vector 
$x \in D \subset \mathbb{R}^{n}$, the diameter D of the orbit is defined as 

\begin{align}\label{Eq:Diameter}
	\textrm{D}(x_{0},y_{0},\tau_{\textrm{run}}) = \max_{0\le t,s \le \tau_{\textrm{run}}} \norm{x(t)-x(s)}.
\end{align}
For our computations we chose the $\ell_{\infty}$-norm and computed the drift along the normalised action variables, \ie along ${\color{black}x=(\tilde{G},\tilde{H})}$ with 
$\tilde{G}=G/L$ and
$\tilde{H}=H/L$ (we recall that in the secular approximation, $L$ is a constant parameter determined by the semi-major axis). 

The results of the computation of the diameters, according to Eq.\,(\ref{Eq:Diameter}) for two-extreme non-linearity parameters are shown in Fig.\,\ref{fig:Diameters}. 
Comparing the results with the FLIs maps, we note that the relation between hyperbolicity and 
large transport is not that straightforward. For $a=18,600$ km, we remark that regions with the 
larger FLIs do not necessarily correspond to regions where the transport is maximal. 
Conversely, the almost vertical resonant manifold emanating near $i \sim 56.1^{\circ}$ does not 
have the largest degree of hyperbolicity; yet it carries the largest transport index. Switching to 
$a=29,600$ km, we note that the lowest diameter is already one order of magnitude larger than 
in the former case. The largest diameter is also significantly larger which confirm the known fact 
of the instabilities in the MEOs.  We emphasise that the diameters have been computed on the 
same predefined grids of ICs used to estimate the FLIs (\ie a highly-resolved grid of ICs). The 
emanating feeling of a resolution deterioration in the maps is once again a nice testimony of the 
sensitivity of variational indicators.

For large perturbing parameters,  globally speaking, large hyperbolicity corresponds {\color{black}to} large 
diameters. This fact has  to be nuanced slightly near $e\sim0.7$ and $i \sim 70^{\circ}$. Using 
an empirical criterion, we extracted from the maps the actions {\color{black}that} satisfy the condition $
\textrm{FLI}(x,\tau_{\textrm{run}}) \ge 1.2 \log (\tau_{\textrm{run}})$ (\ie chaotic orbits) as those 
satisfying $\textrm{D}(x,\tau_{\textrm{run}}) \ge 0.35$. The tracing {\color{black}orbits} are shown in Fig.\,\ref{Fig:FLIvsD} and illustrate the link between large hyperbolicity and large diameters, and the 
necessity of {\color{black}finely} resolved meshes (thin stable structures stripe the chaotic domains and can be 
detected with the diameters also).

Let us now comment on the diameter indicator that we used. 
{\color{black}Very often}
diameters-like quantities in terrestrial dynamics have been estimated using a more 
restrictive definition, namely a one-dimensional diameter of a specific observable 
${\color{black}f}$ \citep[see, \eg][]{eAl16,aRo17}. 
This strategy reduces to nothing else than the amplitude estimation, {\color{black} equivalent to the estimation of $\Delta f = \max_{t} f(x) - \min_{t}f(x)$}. For the 
MEO problem, the eccentricity diameter along the time is, rightly, tracked ({\color{black}as efforts are directed towards the perigee height and the need of re-entry solutions}).
{\color{black}However, when used as 
an empirical `measure of chaos', this diameter may be too loose.} In fact, having in mind the geography of the resonant 
manifolds derived from the resonant condition in Eq.\,(\ref{Eq:ResCond}) and the fact that two 
actions characterises an invariant torus of $\mathbb{T}^{3}$,  it is easy to `create' a quasi first-integral by choosing ICs near certain manifold. 
As an example, let us fix $a=29,600$ km and consider a cluster of ICs in a small neighbourhood of $\mathcal{V}(x_{\star})$, where $x_{\star}=(e_{\star},i_{\star})=(0.616,88^{\circ})$. 
The time evolution (over $25$ lunar periods) of the eccentricity and inclination for the whole 
cluster of orbits ($k=200$ orbits) is displayed in Fig.\,\ref{fig:EnsembleDiameter}.  The spatial 
averaged orbit is displayed and superimposed with a {\it bold red} line\footnote{
Let us consider a cluster of size $k$, that we propagate up to  time $\tau_{\textrm{run}}$. We obtain  $k$ orbits $\gamma_{k} \in \big(\mathcal{C}[0,\tau_{\textrm{run}}]\big)$.  Le us denote by $x_{j}(i,t)$ the instantaneous value of the $j$-th component of the orbit $\gamma_{i}$ at a specific epoch $t$.  The spatial averaged orbit of the component $j$ ($1 \le j \le n$), $\langle x_{j} \rangle$, is then defined through  its components obtained at
any time $t$ by
$
	\langle x_{j}(t) \rangle = \frac{1}{k}\sum_{i=1}^{k} x_{j}(i,t).
$
}.
Clearly, the eccentricities of the whole cluster evolve in an apparent  regular fashion. All the 
orbits incorporate  similar dynamical informations, both  on the quantitative and qualitative point 
of view. On the contrary, the inclination time-histories experience significant variations and a net 
sensitive dependence upon the ICs. 
From this example, easily generalisable, we easily infer why a one-dimensional diameter (based 
on the eccentricity) would fail in capturing these particularities. Pushing further the idea, we 
extended this approach on a grid of ICs near the point $x_{\star}$ by computing accordingly the 
diameters (and the FLIs).  The obtained maps are presented in Fig.\,\ref{fig:D1dvsD2d}. They 
confirm the rationale behind the intuition developed through the former example. Whilst the 
diameter based on both actions is in agreement with the FLI map, the method based on the one-diameter approach give {\color{black}an} irrelevant and {\color{black}uniform} signal. 

Having presented  a general way to quantify the drift, let us focus now on how the drift is mediated in the phase space. 
 
%---------------------------------------------------
\subsection{Visualisation of transport}
%---------------------------------------------------
In the previous sections, we computed FLIs and diameters in various sections 
\begin{align}
	\mathcal{S}(v)=\big\{ (x,y) \in D \times\mathbb{T}^{3} \, \vert \, y=v, v \in \mathbb{T}^{3} \big\}
\end{align}
with $D \subset \mathbb{R}^{2}$.
By fixing $y=0$,  particular   features  in $\mathcal{S}(0)$ have been depicted. In order to visualise transport properties, and to show how its mediation is related to the detected hyperbolic web, we use projection and visualisation techniques 
that have been extensively used over {\color{black}the past} decade to study transport in nearly-integrable Hamiltonian system, symplectic Maps and in Dynamical Astronomy  \citep{mGu02,eLe03,pCi08,riPa15,nGu17}. For a recent overview specifically around the FLIs and their applications, we advise the reader to consult \cite{eLe16} for a pedagogical introductory note. 
The methodology consists in the following. First, we compute the FLIs over a  section $\mathcal{S}(v)$, say on $\mathcal{S}(0)$\footnote{
In this work we were interested in the action-action plane, but the approach can be extended to action-angle or angle-angle planes. For example, a angle-angle section can be defined as
$
	\mathcal{T}=\big\{ (x,y) \in D \times\mathbb{T}^{3} \, \vert \, (y_{1},y_{2}) \in B \subset \mathbb{T}^{2}, x \in D, y_{3}=v_{3} \big\}.
$
}. After this step, we are then able to recognise initial conditions  close to hyperbolic borders or immersed within the chaotic sea. We then select one IC of interest in  $\mathcal{S}(0)$. Let $x_{\star} \in \mathcal{S}(0)$ denotes this IC. Next, we define
a small neighbourhood $\mathcal{V}(x_{\star})$ of $k$ ICs of $x_{\star}$. In theory, it would be sufficient to deal with the sole numerical propagation up to $\tau_{\textrm{run}}$ of the orbit emanating from $x_{\star}$. However,  the procedure is computationally facilitated by considering a cluster of $k$ orbits. 
From these computed orbits $\gamma_{k}(t)  \in \mathcal{C}\big([0,\tau_{\textrm{run}}]\big)$, 
we keep trace only of the points {\color{black}that} return \textit{close enough} to the section $\mathcal{S}(v)$. For that purpose, we introduce 
the family of sections
$\{\mathcal{S}_{\delta}(v)\}_{\delta}$ which are $\delta$-close to $\mathcal{S}(v)$. These sections are defined as
\begin{align}
	\mathcal{S}_{\delta}(v)=\big\{ (x,y)\in D \times\mathbb{T}^{3} \, \vert \, \norm{y - v} \le \delta \big\}, \, \delta \ll 1\in \mathbb{R}_{+}.
\end{align}
When $\delta \to 0$, we recover the `exact' section $\mathcal{S}(v)$.
The introduction of this family of section is essentially to circumvent numerical limitations. Firstly, we deal with a finite time $\tau_{\textrm{run}}$ (that we would like to keep `as small as possible' but `large enough' to extract dynamical mechanisms). Secondly, in practice we do not deal with an orbit 
$\gamma_{k}(t)  \in \mathcal{C}\big([0,\tau_{\textrm{run}}]\big)$, but with a discretised version of this orbit computed, say (to facilitate the exposition), at each  multiple of the fixed step size $\Delta t$, 
$\{\gamma_{k}(t),\, t=i\Delta t\}_{i=0}^{n}$, $n \Delta t = \tau_{\textrm{run}}$. All points of the orbits $\gamma_{k}(t)  \in \mathcal{C}\big([0,\tau_{\textrm{run}}]\big)$ {\color{black}that} return during the simulation to a section of $\{\mathcal{S}_{\delta}(v)\}_{\delta}$ are
identically projected into the exact section $\mathcal{S}(v)$, on which the FLIs are used as a background. By doing that, we are able to relate transport with the web detected by the FLIs.
In our computation we dealt with the $\ell_{\infty}$-norm, $\delta$ is problem dependent and best determined by a calibration procedure\footnote{To give an idea of the size of $\delta$, the results presented in this manuscript have been obtained with $\delta = 0.08$ for the small range of $\varepsilon$, $\delta = 0.1$ for larger range. Different admissible $\delta$ just change the number of points on the section {\color{black}collected}, but leave invariant the transport properties (angles are expressed in radians).}. Finally we worked with a cluster of size $k=200$ initial conditions.

\begin{figure}
\includegraphics[width=0.99\textwidth]{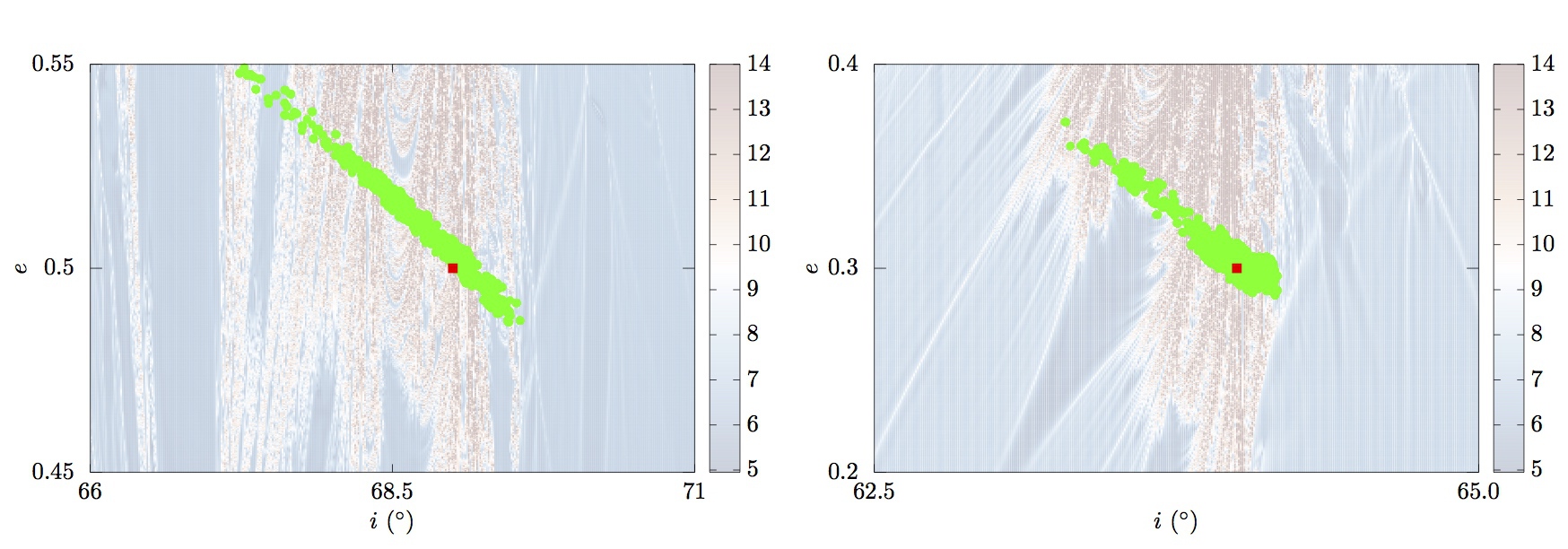} 
\caption{\label{fig:DiffusionSmallEps} 
{\color{black}Two diffusive scenarios illustrating transport scenarios along resonances in the regime of small perturbative parameters. The {\it red} spot indicates the initial condition where the ensemble of initial conditions are defined. Points of the orbits {\color{black}that} return sufficiently close to the section (on which the FLIs appear as a background) are depicted with a {\it green} point. See text for comments.}}
\end{figure}

Fig.\,\ref{fig:DiffusionSmallEps} presents results in the range of 'small' perturbation for {\color{black}two initial} points of interest applying the methodology described previously.  
The clusters has been propagated up to a timescale  of  about $5.8 \times 10^{5}$ orbits revolution ($25$ lunar nodes). The ICs serving a definition to the cluster are depicted in {\color{black}red}. The  points of the orbits {\color{black}that} cross the double sections of the set $\{\mathcal{S}_{\delta}\}_{\delta}$ are depicted in {\it green}. {\color{black}(To facilitate the reading and interpretation of the figures, the FLIs background have been color coded with a light opacity-like filter. The points returning to the section are intentionally magnified.)} The two clusters
focus on thin manifold {\color{black}that} still carry transport (see Fig.\,\ref{fig:Diameters}). {\color{black}As the transport index is rather small, excursions are modest and rather confined.
The returning points are guided by the thin hyperbolic skeleton detected by the FLI computation.}

\begin{figure}
\includegraphics[width=0.99\textwidth]{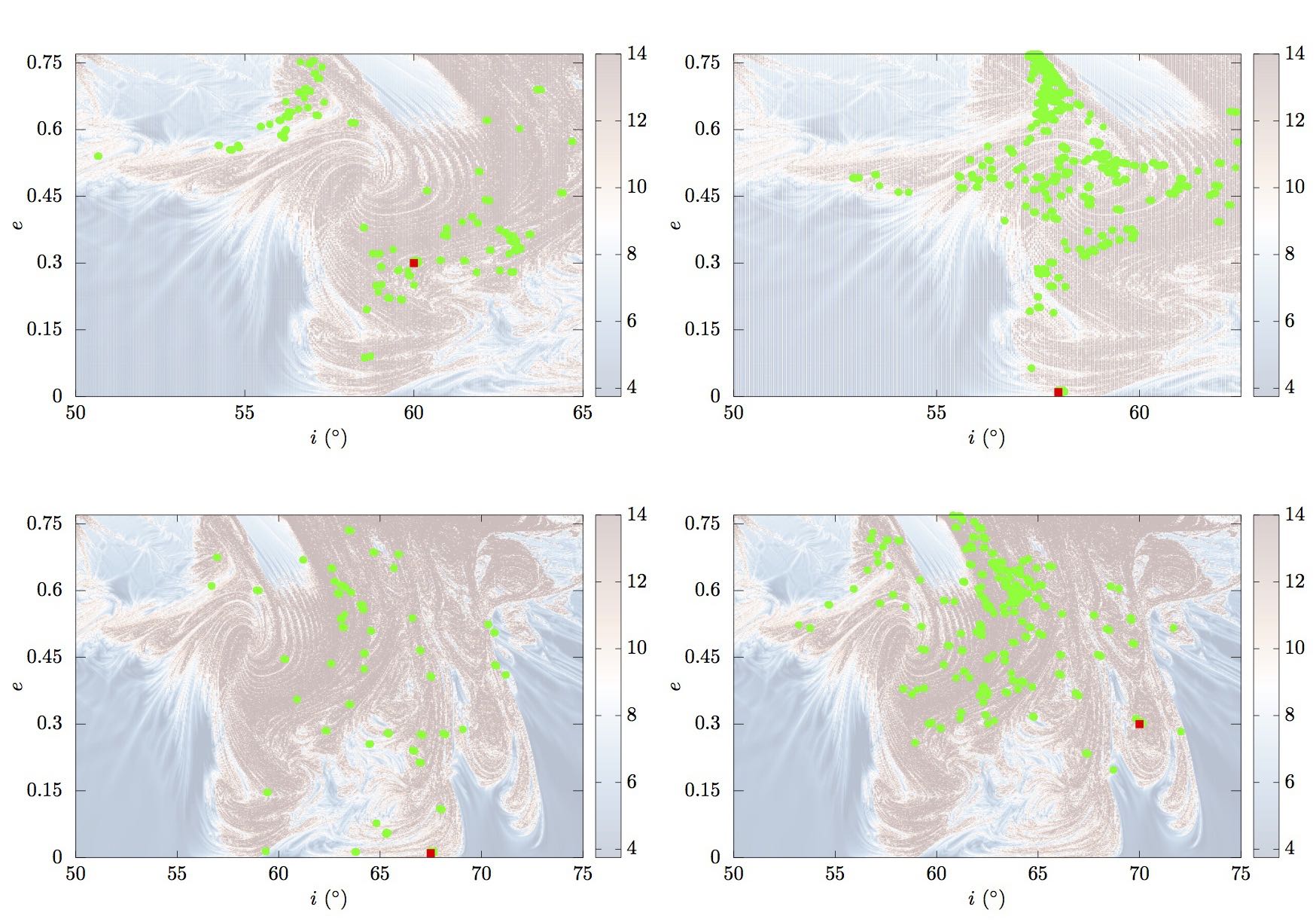} 
\caption{\label{fig:DiffusionLargeEps} 
{\color{black}Diffusive scenarios illustrating transport scenarios within the hyperbolic web in the regime of large perturbative parameters and strong resonances overlap. The {\it red} spots indicate the initial condition where the ensemble of initial conditions are defined. Points of the orbits {\color{black}that} return sufficiently close to the section (on which the FLIs appear as a background) are depicted with a {\it green} point. See text for comments.}}
\end{figure}

In Fig.\,\ref{fig:DiffusionLargeEps} {\color{black}we repeated the same experiment with a larger $\varepsilon$ for $4$ different scenarios}. The approach enables us to visualise and quantify the spread of the actions in the regime of strong chaos. The orbits of the clusters have been propagated on about  $1.4 \times 2.8 \times 10^{5}$ orbits revolution. The spread of the orbits is well more appreciable {\color{black}and develops more drastically within the action-space}. It covers a large portion of the connected chaotic domain.  As it is observed {\color{black} for all scenarios}, the change in inclination can be superior  to $15^{\circ}$, with {\color{black}extremely large} variations for the eccentricity ({\color{black}namely, the mechanism allows nearly circular orbits to become very eccentric}).   

\begin{figure}
\includegraphics[width=0.99\textwidth]{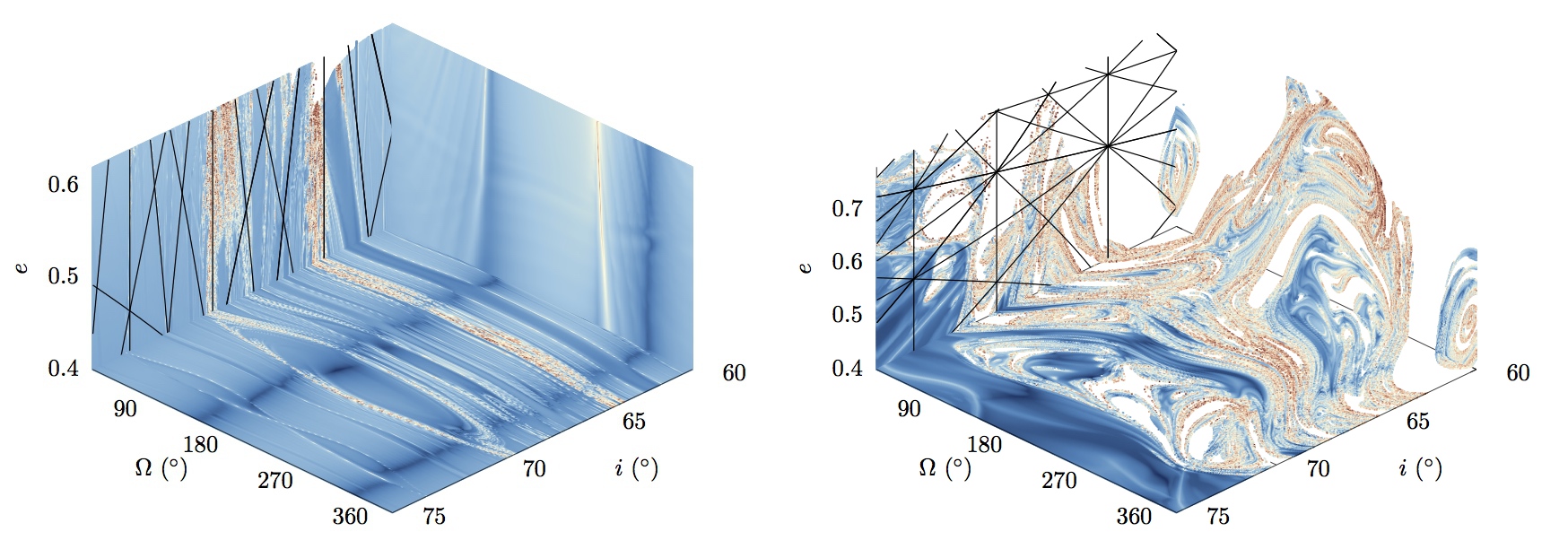}
\caption{\label{fig:Cube} 
{\color{black}These two FLI cubes,  computed for $a=18,600$ km (left) and $a=29,600$ km (right)
highlight a bifurcation phenomena.
The resonant manifolds appear in black in the 'action-action' space.} 
}
\end{figure}

{\color{black}
It would be extremely interesting to extend the approaches and visualisation of the diffusive properties by extending the dimension of the visualised space. Taking advantage of our model, we were able to extend the traditional stability maps in one more direction by stitching together ad-hoc others FLI sections. The results presented in Fig.\,\ref{fig:Cube} complement the global stability picture of the actions space by `unrolling' the dynamics according to one angle, here $\Omega$. The resonant manifolds computed using Eq.\,(\ref{Eq:ResCond}) are depicted in black in the ``action-space''. In the regime of small perturbation (left panel), a pendulum-like structure is clearly identifiable (minor structures can also be identified). By varying the size of the perturbation, a bifurcation-like phenomena occurred and the initially elliptic point becomes of a hyperbolic nature where collisional orbits develop. Such a systematic parametric methodology would allow,  besides the quantification of chaos and the determination of the resonant regime \citep[cf.][]{cFr00}, the determination of precise perturbing parameters where such phenomena occur.}

%----------------------------------------------------
\section{Discussion and conclusive remarks}\label{sec:discussion}
%----------------------------------------------------
Dynamical chaos indicators as the FLI are valuable and formidable allies to gain knowledge on the dynamical system under investigation.  Their systematic use over nearly the past two decades in transverse fields has brought its share of results. Applications towards terrestrial dynamics are still at their early stage but the current situation seems to evolve positively.  
In this contribution, we complemented and refined our past studies related to the long-term dynamics of terrestrial orbits in the range $2.91$ to $4.64$ Earth radii ($\varepsilon \in [0.02:0.22]$). We showed the complementarity and benefits of visualising the global dynamics via sections, corroborated with the  computation of the FLIs and practical action-diameter quantities. From our numerical experiments, we have seen that when the detected hyperbolic manifolds are very thin (but still carry large diameters), the transport occurs precisely along them. For higher values of the non-linearity parameter, resonances do overlap significantly  and the transport is across a large domain of the chaotic sea. This mechanism allows nearly-circular orbits to become highly eccentric on a few lunar nodes only.  In the later case of strong chaos, preferred directions for the transport are hard to establish. The FLIs allow to follow and delineate the routes of transport where the spread in the phase space take place.
The natural  complementary step that deserves serious attention concerns  the nature of the transport, the computation of \textit{diffusion-{\color{black}like} coefficients} and its scaling with $\varepsilon$. (Note that if in our actual set-up, we do have access only to a limited number of different order of magnitudes of $\varepsilon$. A theoretical possibility to extend its range is to artificially increase the semi-major axis - even if we know that physically the procedure is not that relevant as octupolar contributions should be incorporated.) Let us comment and relate recent difficulties that we encountered in investigating these last points. 
Transport properties are generally characterised through the computations of moments of different order $q$,
\begin{align}
	M_{q}(\tau) = \big\langle \vert x(\tau) - \langle x(\tau) \rangle \vert^{q} \big\rangle.
\end{align}
Let us underline once more that when we deal with the dynamics numerically, we only have access to \textit{finite time moments}. Usually, the second-order moment, \ie the spread of the actions (the variance), is used to discriminate the case of diffusion we deal with. 
More precisely, under the explicit ansatz that
\begin{align}
	M_{2}(\tau) = \big\langle \vert x(\tau) - \langle x(\tau) \rangle \vert^{2} \big \rangle \sim D_{2} \tau^{\nu},
\end{align}
the diffusion is called either \textit{subdiffusive} ($\nu < 1$), \textit{diffusive} ($\nu = 1$) or \textit{superdiffusive} ($\nu > 1$). (The particular case of superdiffusive behaviour with $\nu=2$ is referred to \textit{ballistic diffusion}.) The real parameter $D_{2}$ is the estimated \textit{diffusion coefficient}, and its sole  determination can be sometimes tricky due to technical difficulties (see, \eg \citet{eLe03} and further references in \citet{pCi18}). Anomalies to the strict diffusive case ($\nu=1$), \ie aberrations with respect to Gaussianity, might be the results of the existence of a mixed phase space (cohabitation of regular and chaotic components in the phase space) and correlation effects \citep{hVa97,gZa02}. Let us note that, to the best of our knowledge, the study of the correlation function $C(\tau)$ (even at least for the  specific observable of interest, the eccentricity) and its possible decay which give us the scale of the \textit{correlation time} $\tau_{C}$ \citep[see discussions in][]{sWi04,hVa05} has never been undertaken for the MEO problem. {\color{black}(The exception is found in \cite{iWy07}  where the autocorrelation function properties are used to discriminate regularity for geosynchronous objects.)}
\cite{jDa17} claimed the normal character of the diffusion for the eccentricity observable in the regime of strong chaos. We redid 
some experiments along those lines apart that we used the spatial averaging ideology (and no longer the \textit{temporal averaging}) assuming that all ICs of the cluster are equivalent. We met difficulties to confirm our former conclusions and we stress here that they should be taken with a grain of salt.  
In fact, in our experiments, we noticed that such a conclusion depends \textit{strongly} on the ansatz made on the evolution of the variance and  the choice of the time-horizon investigated. Regarding the question related to the time-horizon, there might exist a transient time $\tau_{\textrm{tr.}}$ that should be constrained first. Indeed, in order to derive meaningful statistical conclusions, we have to ensure that $\tau \gg \tau_{\textrm{tr.}}$ (as a transient time seems to exist) and
$\tau \gg \tau_{C}$. It is possible that, unfortunately, in our present setting, $\tau_{\textrm{tr.}} \sim \tau$, making conclusions hard to reach. 

Constraining those difficulties are the directions being taken by our current research.

\section*{Author Contributions}
The paper has been written by the authors in equal parts.

\section*{Acknowledgments}
The authors greatly appreciated and acknowledge the discussions with Christos Efthymiopolous at the `Perspectives in Hamiltonian dynamics' conference held in Venice, June $18-22$. J.D.\, is grateful to Bastian M\"arkisch of the SourceForge \texttt{gnuplot} forum for his precious advices  in realising Figs.\,\ref{Fig:FenceProgradeFLIs-part1} and \ref{Fig:FenceProgradeFLIs-part2} and thank Beno\^it Noyelles for his remarks.  Numerical simulations were performed using HPC resources from the Institute for Celestial Mechanics and Computation of Ephemerides (IMCCE, Paris Observatory) and computing facilities from RMIT University. J.D.\, acknowledges the support of the Cooperative Research Centre for Space Environment Research Centre (SERC Limited) through the Australian Government's Cooperative Research Center Program and the support of the ERC project $677793$ `Stable and Chaotic Motions in the Planetary Problem'. I.G.\,acknowledges the support of the ERC project 679086 `Control for Orbit Manoeuvring through Perturbations for Application to Space Systems'.

\bibliographystyle{apalike}
\bibliography{biblio}

\end{document}